\documentclass[10pt,journal]{IEEEtran}

\usepackage{graphicx,amssymb,lineno}
\usepackage{amsmath,amsfonts,amssymb}

\usepackage[usenames]{color}
\usepackage{float}
\usepackage{epstopdf}
\usepackage{longtable,booktabs}
\usepackage[numbers,sort&compress]{natbib}
\usepackage{mathtools}
\usepackage[justification=centering]{caption}

\makeatletter
\newif\if@restonecol
\makeatother

\usepackage[linesnumbered,ruled,vlined]{algorithm2e}
\usepackage{algpseudocode}

\usepackage{chngpage}
\usepackage{array}

\usepackage{multirow}
\usepackage{graphicx,graphics,color,epsfig,subfigure,graphpap,rotate}
\usepackage{times, verbatim, subfigure, epsfig, graphicx, latexsym, amsmath}
\usepackage{url}
\usepackage{subfigure}

\begin{document}
\setlength{\baselineskip}{11.5pt}
\title{Triple-Band Scheduling with Millimeter Wave and Terahertz Bands for Wireless Backhaul  }

\author{Yibing~Wang,
        Hao~Wu,
        Yong~Niu,~\IEEEmembership{Member,~IEEE},
        Jianwen~Ding,
        Shiwen~Mao,~\IEEEmembership{Fellow,~IEEE},
        Bo~Ai,~\IEEEmembership{Fellow,~IEEE},
        Zhangdui~Zhong,~\IEEEmembership{Fellow,~IEEE},
        and Ning~Wang
\thanks{Y. Wang, H. Wu, J. Ding, B. Ai and Z. Zhong are with the State Key Laboratory of Rail Traffic
 Control and Safety, Beijing Jiaotong University, Beijing 100044, China (E-mails:
 18111034@bjtu.edu.cn, hwu@bjtu.edu.cn, jwding@bjtu.edu.cn, boai@bjtu.edu.cn, and zhdzhong@bjtu.edu.cn).} 
\thanks{Y. Niu is with the State Key Laboratory of Rail Traffic Control and Safety,
Beijing Jiaotong University, Beijing 100044, China, and also with the National
Mobile Communications Research Laboratory, Southeast University, Nanjing
211189, China (Email:niuy11@163.com).}
\thanks{S. Mao
is with the Electrical and Computer Engineering Department, Auburn University, Auburn, AL 36849-5201, USA (Email: smao@ieee.org). }
\thanks{N.~Wang
is with the School of Information Engineering,
Zhengzhou University, Zhengzhou 450001, China (Email: ienwang@zzu.edu.cn). }
}%

\maketitle

\begin{abstract}
With the explosive growth of mobile traffic demand, densely deployed
small cells underlying macrocells have great potential for 5G and beyond wireless networks.
In this paper, we consider
the problem of supporting traffic flows with diverse QoS requirements by exploiting three high frequency bands, i.e., the 28GHz band, the E-band, and the Terahertz (THz) band.
The cooperation of the three bands
is helpful for maximizing the number of flows
with their QoS requirements satisfied.
To solve the formulated
nonlinear integer programming problem,
we propose a triple-band scheduling scheme which can select the optimum scheduling band for each flow among three different frequency bands.
The proposed scheme also efficiently utilizes the resource to schedule flow transmissions in time slots.
Extensive simulations
demonstrate the superior performance of the proposed scheme over three baseline schemes with respect to the number of completed flows and the system throughput.
\end{abstract}


\section{Introduction}\label{S1}

Over the last few years, the explosive growth of mobile data demand has always been the focus of attention. Edholm's law reveals that wireless data rates have doubled every 18 months. Moreover, the world monthly mobile traffic will be about 49 exabytes by the
year 2021~\cite{growth}.
It is apparent that the existing communication technologies and the available spectrum resources cannot fully support such huge data demand. Therefore, many solutions have been proposed to
deal with the problem of excessive data demand.
Among the many potential solutions, densely deployed
small cells underlying macrocells
are shown effective on achieving
higher network capacity~\cite{cell,Zhou15}.
Traffic offloading with small cells can effectively ease the
burden on macrocells~\cite{s-cell}. The short transmission range
in small cells enhances the
transmission quality and leads to higher transmission rates. Although various technologies (such as modulation, multiplexing, multiple antennas, etc.) have been used to deal with the massive traffic demands,
the communications in higher frequency bands
(i.e., spectrum expansion)
seems to be the most effective means to deal with
the ever-increasing rate demands~\cite{higher}.
To this end, millimeter-wave (mmWave) communications have been proposed as a key component of 5G wireless networks~\cite{mmWave}. However, substantial studies have shown
the limits of the
current wireless networks that rely on the mmWave frequency bands. In order to obtain greater transmission bandwidth, Terahertz (THz) band should be utilized to even further increase the achievable rates of
the 5G beyond wireless networks.
For wireless communications in the
THz band, apart from issues of electronics (such as the design
of high power compact transceivers~\cite{manufacture}), the very
high free-space attenuation by spreading loss is also a big challenge.

In fact, behind the explosive growth of data demand is a large number of
existing and emerging applications.
The transmission rate requirements of different applications are highly diverse.
For example, web browsing usually only requires a multi-Mbps transmission rate, but
streaming an
uncompressed high-definition TV (HDTV) requires a
multi-Gbps rate~\cite{HD,HD1,He14}. Although the single high frequency band with enough bandwidth may
be used for the scheduling of a large number of different flows,
the advantages would be obvious if multiple bands can be utilized to schedule the diverse traffic flows.
According to the transmission characteristics of different frequency bands, choosing the appropriate band for different flow transmissions is conducive to improving the system performance.

In this paper, we consider the backhaul network with
densely deployed small cells underlying a macrocell.
There are a set of flows to be scheduled with their minimum throughput requirements (i.e., the lowest transmission rate requirements), which are also referred to as the quality of service (QoS) requirements in this paper. QoS requirements of different flows are highly diverse.
We propose to exploit two mmWave bands (including one lower frequency mmWave band
and the E-band) and one THz band
to carry these flows. The proposed band selection algorithm determines the transmission band of each flow based on its QoS requirement and transmission range.
Moreover, we propose a triple-band scheduling algorithm to
schedule transmissions of the flows in different bands and time slots. The contributions of this paper can be summarized as follows.
\begin{itemize}
  \item
  To fully exploit the available spectrum resources, we propose the cooperation of three bands (i.e., two mmWave bands and one THz band) to schedule a large number of flows with diverse
  QoS requirements. In order to avoid the pressure
  putting all flows in the highest frequency band, we disperse flows with relatively
  lower QoS requirements to the lower frequency band for transmission. This approach allows more resources for other flows with
  more stringent QoS requirements.
  \item We formulate the problem of optimal triple-band scheduling as a nonlinear integer programming problem. Moreover, we propose a heuristic triple-band scheduling algorithm to maximize the number of flows with their QoS requirements satisfied
  within a fixed time. The algorithm determines whether flows conflict
  with each other
  based on their
  mutual interference.
  \item We evaluate the proposed triple-band scheduling scheme for the backhaul network in the 28GHz, 73GHz, and 340GHz bands with extensive simulations. The simulation results demonstrate that our proposed scheme
  outperforms three baseline schemes
  on both
  the number of completed flows and the system throughput.
\end{itemize}

The remainder
of this paper is organized as follows. Section~\ref{S2}
reviews the related work. Section~\ref{S3} introduces the system model. In Section~\ref{S4}, we formulate the optimal triple-band scheduling problem as a nonlinear integer programming problem. In Section~\ref{S5}, we propose the triple-band scheduling scheme with the transmission band selection algorithm. In Section~\ref{S6}, we evaluate the proposed scheme
with simulations. Section~\ref{S7}
concludes this
paper.

\section{Related Work}\label{S2}

There are many studies on flow scheduling in mmWave networks with a single frequency band. In~\cite{TDMA}, a scheduling scheme based on time division multiple access (TDMA) is proposed to support the communications from one point to multiple points in mmWave backhaul networks. In~\cite{add1}, the authors design a scheduling algorithm for backhaul networks by exploiting mmWave macro base stations as relay nodes. In~\cite{add2}, an efficient polynomial-time scheduling method and an approximation algorithm parallel data stream scheduling method are proposed for no interference network model. In~\cite{add3}, the authors optimize the scheduling of access and backhaul links to maximize the achievable minimum throughput of the access links. In~\cite{add4}, an algorithm is proposed to find high throughput paths with relays for links by minimizing interference within and between paths.  In~\cite{problem}, a relay selection algorithm and a transmission scheduling algorithm are proposed to relay the blocked flows and maximize the number of completed flows. In this paper, we consider the transmissions in the THz band in addition to the mmWave band, and aim to maximize the number of scheduling flows.

In addition, there are also
some related works on communications in the THz band. In~\cite{LOS},
the authors propose an approach to enhance the performance of cellular networks with THz links.
In~\cite{THZ}, the authors propose an algorithm of QoS-aware bandwidth allocation and concurrent scheduling. In~\cite{THZ1},
a sub-channel allocation method and power allocation scheme based on improved Whale Optimization Algorithm is proposed.
Furthermore, multi-band cooperation has also been applied in a few areas. In~\cite{growth}, a control-data separation architecture for dual-band mmWave networks is proposed. In~\cite{dual}, a wireless local area network (WLAN) architecture
utilizing the new multi-beam transmissions in sub-6GHz and mmWave dual-band cooperation mechanisms is proposed to improve the throughput and reliability of the network. This paper variously seeks a mode of cooperation in mmWave and THz band, instead of only focusing on mmWave communication or THz communication as the works listed above.

However, these existing related works have not considered the case of a triple-band cooperation involving both mmWave and THz communications. To the best of our knowledge, there is no research on a scheduling problem that incorporates multi-band transmissions. This is mainly due to the complexity of THz transmission itself and the high cost of transmission equipment.
In this paper, through the
integration of multi-band cooperation into the scheduling problem,
we will show that
the performance with respect to the number of completed flows and the system throughput
could be greatly improved.

\section{System Overview}\label{S3}

\begin{table}[!t]
  \centering
  \caption{Summary of Notation and Description}\label{Summary}
\begin{adjustwidth}{0.1cm}{0.5cm}
 \begin{tabular}{|p{.15\textwidth}| p{.25\textwidth}| m{.3\textwidth}|}
 \hline
Notation & Description \\
\hline
$M$, $\Delta t$, $t_0$ & number of TSs, time duration of the TS, time duration of the scheduling phase \\
\hline
${P_r^{mmWave}}(i,i)$, ${P_r^{THz}}(i^\prime,i^\prime)$ & received power of flow $i$ in mmWave band (28GHz band or E-band), received power of flow $i^\prime$ in THz band\\
\hline
$P_t^{mmWave}$, $P_t^{THz}$ & transmit power of mmWave transmission (28GHz transmission or E-band transmission), transmit power of THz transmission \\
\hline
$G^t_{s_i}$, $G^r_{r_i}$ & directional antenna gain at transmitter $s_i$, directional antenna gain at receiver $r_i$ \\
\hline
$d_{s_ir_i}$ & distance between $t_i$ and $r_i$ \\
\hline
$f^{THz}$ & carrier frequency of THz band\\
\hline
$\lambda^{mmWave}$, $\lambda^{THz}$ &  wavelength of mmWave band (28GHz band or E-band), wavelength of THz band \\
\hline
$k_0^{mmWave}$ &  free space path loss of mmWave transmission\\
\hline
${I_{ji}^{mmWave}}$, ${I_{{j^\prime}{i^\prime}}^{THz}}$ &  interference at $r_i$ from $s_j$ in mmWave band, interference at $r^\prime_i$ from $s^\prime_j$ in THz band \\
\hline
${l^{THz}}(i^\prime,i^\prime)$ &  path loss of the flow $i^\prime$ in THz band \\
\hline
$R_i^{mmWave}$, $W_i^{mmWave}$ &  transmission rate of flow $i$ in mmWave band (28GHz band or E-band), bandwidth of mmWave band (28GHz band or E-band)\\
\hline
$W^{mm}$, $W^{me}$, $W^{THz}$ &   bandwidth of 28GHz band,  bandwidth of E-band, bandwidth of THz band \\
\hline
$N_0^{THz}(i^\prime)$ &  total noise at the receiver of flow $i^\prime$ in THz band \\
\hline
$R_i^{mm}$, $R_i^{me}$, $R_{i}^{THz}$ &  transmission rate of flow $i$ in 28GHz band, transmission rate of flow $i$ in E-band, transmission rate of flow $i$ in THz band \\
\hline
$\theta^t_{ml}$, $\theta^r_{ml}$ &  beamwidth of the main lobe of transmitter antenna, beamwidth of the main lobe of receiver antenna\\
\hline
$q_i$, $T_i$ &  QoS requirement of flow $i$,  actual throughput of flow $i$\\
\hline
$D_{ref}^{THz}$ &  reference distance of THz communication\\
\hline
$\delta$ &  type of bands ($\delta=mm$ is 28GHz band, $\delta=me$ is E-band or $\delta=THz$ is THz band)\\
\hline
$Q_i$ &   binary variable of flow state ($Q_i=0$ indicates flow $i$ is uncompleted or $Q_i=1$ indicates flow $i$ is completed)\\
\hline
$a_{i_{mm}}^t$, $a_{i_{me}}^t$, $a_{i_{T}}^t$ &   binary variable of flow state in 28GHz band in TS $t$ ($a_{i_{mm}}^t=0$ indicates flow $i$ isn't scheduled in 28GHz band in TS $t$ or $a_{i_{mm}}^t=1$ indicates flow $i$ is scheduled in 28GHz band in TS $t$ ), binary variable of flow state in E-band in TS $t$, binary variable of flow state in THz band in TS $t$\\
\hline
 \end{tabular}
\end{adjustwidth}
\end{table}

\begin{figure}[t!]
	\begin{center}
		\includegraphics[width=3.3in]{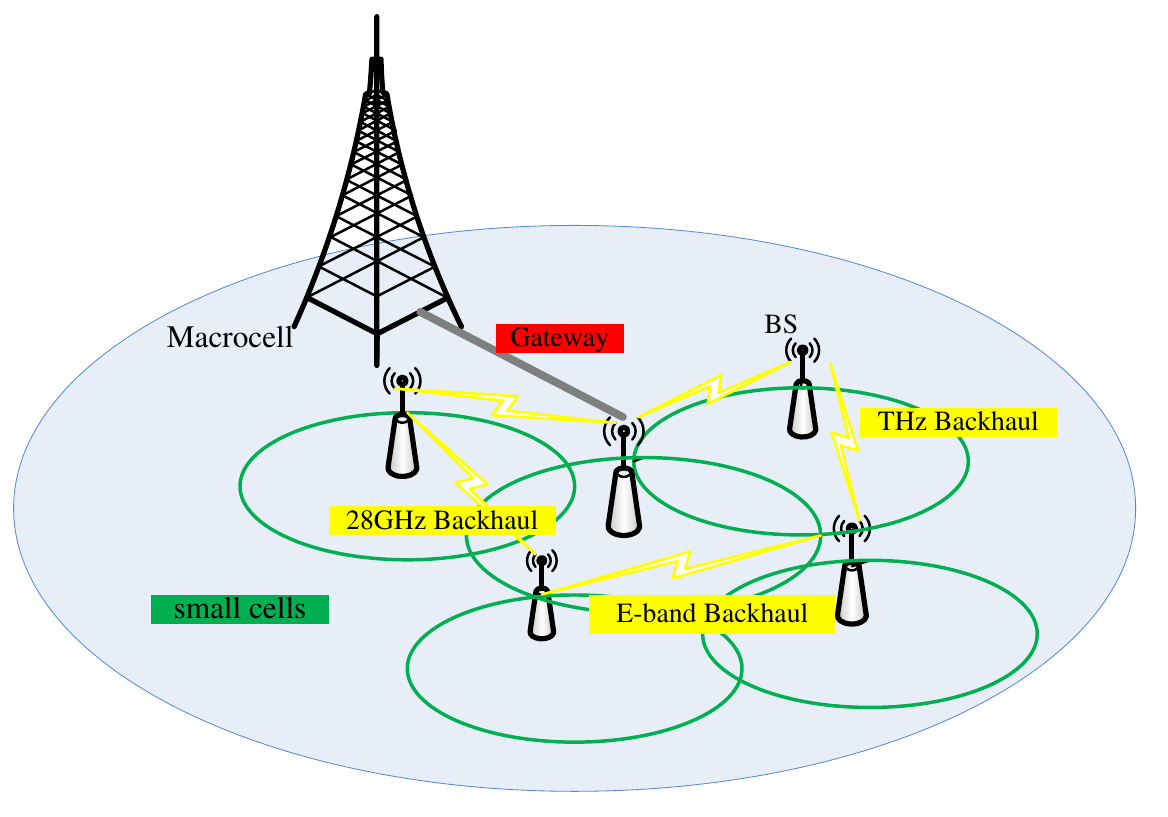}
		\caption{The backhaul network in the small cells densely deployed scenario.}
		\label{fig:backhaul}
	\end{center}
\end{figure}

In this paper, we consider a backhaul network supporting densely deployed small cells,
as shown in Fig.~\ref{fig:backhaul}. The backhaul network controller (BNC) resides on one of the gateways to synchronize the network, receive the QoS requirements of different flows, and obtain the BS locations. Each BS is equipped with an electronically steerable directional antenna which can
transmit in narrow beams towards other BSs, and operate in the half duplex (HD) mode.
If there is a certain amount of traffic demand between any two BSs,
a flow is requested between the pair of BSs.
Each flow has its minimum throughput requirement, which is the QoS requirement of the flow considered
in this paper. To serve a large amount of traffic demand with
a wide
range of QoS requirements,
we adopt a triple-band transmission approach that utilizing the 28GHz band, the E-band, and the THz band.
Compared with the mmWave band, the THz band has
much more bandwidth available, but its higher propagation loss results in shorter transmission ranges.
It is an interesting problem to investigate how to schedule the flows over the multiple bands to maximizing the throughput of the network.

\subsection{System Model}\label{S3-1}

Since non-line-of-sight (NLOS) transmissions suffer from much higher attenuation than line-of-sight (LOS) transmissions
in both mmWave and THz bands~\cite{attenuation}, we assume there is a directional LOS link between any pair of BSs by appropriate adjusting the locations of the BSs.\footnote{Although the communications between a transmitter and receiver
in the THz band can also be made in NLOS
by building reflections,
as the case in the mmWave band~\cite{higher}, due to the lack of corresponding measurement studies of THz communications, NLOS transmissions in mmWave and THz bands are both not considered in this paper.}
In the heterogeneous network, there are $N$ BSs and $F$ flows which need to be transmitted among the
$N$ BSs. The flows can transmit in either one of the mmWave bands or the THz band in this paper. Therefore, we need to consider the transmission models of
a flow in different bands.

In the system, time is divided into a series of non-overlapping superframes, and each superframe consists of a scheduling phase and a transmission phase. The scheduling phase is the time duration to collect
flow requests and their QoS requirements.
The transmission phase consists of $M$ equal time slots (TSs). For clarity of illustration, we use TSs to measure the transmission time. For reliable transmission, the control signaling and BS requests can be collected by the BSs using 4G
transmissions,
when the BNC receives
the source/destination BSs information and QoS requirements for each flow~\cite{4G} (assuming the BS locations are already known, since they are fixed).
The proposed scheduler is to allow multiple flows be transmitted concurrently
using the spatial-time division multiple access (STDMA) in this paper~\cite{greedy}.

\smallskip
\subsubsection{mmWave Transmission Model}\label{S3-1-1}

For a flow $i$ that is
transmitted from transmitter $s_i$
in the
28GHz band or E-band, the received power at its receiver $r_i$
can be calculated as
\begin{equation}\label{pm}
{P_r^{mmWave}}(i,i) = {k_0^{mmWave}}G^t_{s_i}G^r_{r_i}d_{s_ir_i}^{ - n}{P_t^{mmWave}},
\end{equation}
where $k_0^{mmWave}$ is the free space path loss of mmWave transmission
at the reference
distance 1m,
which is proportional to $\left(\frac{\lambda^{mmWave} }{{4\pi }}\right)^2$, where $\lambda^{mmWave}$ is the wavelength
and $P_t^{mmWave}$ is the transmit power of the mmWave transmission~\cite{Pt}; $G^t_{s_i}$ and $G^r_{r_i}$ are the directional antenna gain at $s_i$ and $r_i$, respectively;
$d_{s_ir_i}$ denotes the distance between $t_i$ and $r_i$; and $n$ is the path loss exponent. Since
the BSs operate in the
HD mode, adjacent co-band flows that involve the same BS cannot be scheduled to transmit concurrently. For co-band flows $i$ and $j$
that do not involve the same BS,
the interference at $r_i$ from $s_j$ can be calculated as
\begin{equation}\label{im}
{I_{ji}^{mmWave}} = \rho{k_0^{mmWave}}G^t_{s_j}G^r_{r_i}d_{s_jr_i}^{ - n}{P_t^{mmWave}},
\end{equation}
where $\rho$ is the factor of multi-user interference (MUI) that
is related to the cross correlation of signals from different flows~\cite{MQIS}.

The transmission rate of flow $i$ can be obtained
according to Shannon's channel capacity,
given by
\begin{equation}\label{rm}
\begin{aligned}
&R_i^{mmWave}=\\
&\eta W_i^{mmWave}{\rm{log}}_2\left(1+\frac{{P_r^{mmWave}}(i,i)}
{N_0W_i^{mmWave}+\sum\limits_{j}I_{ji}^{mmWave}}\right),
\end{aligned}
\end{equation}
where $\eta$ is in the range of $(0,1)$ and describes the efficiency of the transceiver design. $W_i^{mmWave}$ is the transmission bandwidth of the flow $i$, and $N_0$ is the onesided power spectra density of white Gaussian noise~\cite{Niu}.

\smallskip
\subsubsection{THz Transmission Model}\label{S3-1-2}

In the case that flow $i^\prime$
is
transmitted
in the
THz band, the path loss is given by
\begin{equation}\label{lt}
{l^{THz}}(i^\prime,i^\prime) = 92.4+20lg(f^{THz})+20lg(d_{s_{i^\prime}r_{i^\prime}}),
\end{equation}
the path loss model is proposed in Ref. \cite{gra}, and is related to the frequency and distance of the transmission~\cite{absorption1, absorption2}. The unit of $f^{THz}$ is $GHz$, and the unit of $d_{s_{i^\prime}r_{i^\prime}}$ is $km$. For the main molecule of the atmosphere, i.e., water-based absorption, the attenuation of some high THz bands can
be as high as hundreds dB/km, and the attenuation of some lower spectrum THz bands
is only few dB/km.

For flow $i^\prime$, the received power at its receiver $r_{i^\prime}$ from its transmitter $t_{i^\prime}$ can be calculated as
\begin{equation}\label{pt}
{P_r^{THz}}(i^\prime,i^\prime) = {l^{THz}}(i^\prime,i^\prime)G^t_{s_{i^\prime}}G^r_{r_{i^\prime}}{P_t^{THz}},
\end{equation}
where $P_t^{THz}$ is the transmit power of the THz transmissions.

Similar to the mmWave case,
we also consider the interference between
co-band concurrent flows that do not involve the same BS (i.e., non-adjacent) in the THz band. For flows $i^\prime$ and $j^\prime$ in THz band, which are non-adjacent, the interference at $r_{i^\prime}$ from $s_{j^\prime}$ can be calculated as
\begin{equation}\label{it}
{I_{{j^\prime}{i^\prime}}^{THz}} = \rho^\prime{l^{THz}}(j^\prime,i^\prime)G^t_{s_{j^\prime}}G^r_{r_{i^\prime}}{P_t^{THz}},
\end{equation}
where $\rho^\prime$ is the factor of multi-user interference (MUI), as the $\rho$ in~(\ref{im}). ${l^{THz}}(j^\prime,i^\prime)$ is the path loss of the flow from $s_{j^\prime}$ to $r_{i^\prime}$ in the THz band, which is calculated as~(\ref{lt}).

Hence, according to Shannon's channel capacity, the transmission rate of flow $i^\prime$ can be calculated as
\begin{equation}\label{rt}
R_{i^\prime}^{THz}=\eta W^{THz}{\rm{log}}_2\left(1+\frac{{P_r^{THz}}(i^\prime,i^\prime)}{N_0 W^{THz}+
\sum\limits_{j^\prime}I_{j^\prime i^\prime}^{THz}}\right),
\end{equation}
where $W^{THz}$ is the bandwidth of the THz band.

There is a tradeoff in the
choice of
the carrier-frequency of the THz transmission, $f^{THz}$,
which
affects both the bandwidth and the path loss of the
THz transmission. The higher the bandwidth that can be utilized,
the higher the
molecular
absorption \cite{absorption1}.
In this paper, we choose
$f_{THz} =340$ GHz. Experiments show that the group velocity dispersion of the channel in this frequency band is very small, and the signal is not easy to be broadened. The maximum transmission rate of this frequency band can achieve more than 10 Gbps \cite{gra}.

The output power of the THz amplifier
is closely related to the electronic-circuit design
and is limited by the current
hardware technology.
For instance, the carrier frequency of 340GHz is adopted in this paper, and the amplified power can achieve around 20mW with the current electronic technology~\cite{absorption1}. Therefore, compared with the transmit power of mmWave transmissions,
the transmit power
of THz transmission is relatively
low, i.e., $P_t^{THz}<P_t^{mmWave}$.

\smallskip
\subsubsection{Antenna Model}\label{S3-1-3}

Since the wavelength of the 28GHz band,
E-band,
and THz band transmissions are all short, we adopt
directional beamforming
by utilizing the large
antenna arrays in
the small cell BSs.
With the directional beamforming technique, the transmitter and the receiver of each flow are able to direct beams towards each other for the directional communication~\cite{beamforming}.

For the 28GHz band and E-band, the gain of the directional antenna is given by a sectored radiation pattern as~\cite{antenna}
\begin{equation}\label{equation:Antenna Model}
G^d(\theta)=\left\{
\begin{array}{ll}
G^d_{max}, & \mid\theta^d\mid\leq\theta^d_{ml} \\
G^d_{min}, & \mid\theta^d\mid>\theta^d_{ml} ,
\end{array}
\right.
\end{equation}
where $d\in\{t,r\}$, $\theta^d$ is the angle of antenna boresight direction, $\theta^d\in[-\pi, \pi)$, and $\theta^d_{ml}$ is the beamwidth of the main lobe, $G^d_{max}$ is the maximum antenna gain of the desired link with the perfect beam alignment, and $G^d_{min}$ is the minimum antenna gain. We assume that the antennas of the transmitter and receiver point towards each other before transmission with the maximum directivity gain $G^t_{max}G^r_{max}$. The beams of
other
interfering flows are
assumed to be
uniformly distributed in $[-\pi, \pi)$, and then there are four possible value of the directivity gain between an interferer and the receiver of interest. The four specific gain values and their respective probabilities are given in Table~\ref{tab:gain}, where $\frac{\theta_t}{2\pi}$ and $\frac{\theta_r}{2\pi}$ are the probabilities that the antenna gains for the alignment directions of an interfering transmitter and the receiver of interest are equal to $G^t_{max}$ and $G^r_{max}$, respectively.

\begin{table}[!t]
\captionsetup{font={small}}
 \caption{\label{tab:gain} Directivity gains of the interfering flows and  probabilities of occurrence}
 \centering
 \begin{tabular}{ll}
  \toprule
  Gain value & Probability\\
  \midrule
  $G^t_{max}G^r_{max}$ & $(\frac{\theta_t}{2\pi})(\frac{\theta_r}{2\pi})$ \\
  $G^t_{max}G^r_{min}$ & $(\frac{\theta_t}{2\pi})(1-\frac{\theta_r}{2\pi})$ \\
  $G^t_{min}G^r_{max}$ & $(1-\frac{\theta_t}{2\pi})(\frac{\theta_r}{2\pi})$ \\
  $G^t_{min}G^r_{min}$ & $(1-\frac{\theta_t}{2\pi})(1-\frac{\theta_r}{2\pi})$ \\
  \bottomrule
 \end{tabular}
\end{table}

For the THz band, the narrow beam antenna model of F. 699-7 recommended by ITU-R is adopted \cite{tan}. The gain relative to the isotropic antenna $G(\varphi)$ is given by:
\begin{equation}\label{equation:THz Antenna}
G(\varphi)=\left\{\begin{array}{ll}
G_{max}-2.5\times 10^{-3(\frac{D}{\lambda^{THz}}\varphi)^2},&0^\circ<\varphi<\varphi_m\\
G_1,&\varphi_m\leq\varphi<\varphi_r\\
32-25log\varphi, &\varphi_r\leq\varphi<48^\circ\\
-13, &48^\circ\leq\varphi<180^\circ,
\end{array}\right.
\end{equation}
where $\varphi$ is the off-axis angle; $G_{max}$ is the maximum antenna gain, which is the antenna gain of the main lobe;
$D$ is the antenna diameter; $\lambda^{THz}$ is the wavelength of THz band;
$G_1=2+15log(\frac{D}{\lambda^{THz}})$ is the gain of the second side lobe; $\varphi_m=\frac{20\lambda^{THz}}{D}\sqrt{G_{max}-G_1}$; and $\varphi_r=15.85(\frac{D}{\lambda^{THz}})^{(-0.6)}$. When $G_{max}=47$dBi and $\frac{D}{\lambda^{THz}}=152$, the antenna is the directional antenna, which is called the cassegreen antenna. The unit of the gains in this model is dBi.
This antenna model
is suitable for communication systems in the relatively low THz band.
\section{Problem Formulation}\label{S4}

We present the problem formulation in this section.
For the above network and triple-band transmission model,
the goal of the scheduling
scheme is to accommodate as many flows as possible in the limited time slots. A binary variable $Q_i$ is defined to indicate whether a flow $i$ achieves its QoS requirement under the current transmission schedule.
We have $Q_i=1$ if that is the case, and $Q_i=0$ otherwise.
Then the objective function of the scheduling algorithm can be formulated as
\begin{equation}\label{objective}
\text{max} \sum_{i=1}^F Q_i.
\end{equation}

For flow $i$, we define three binary variables $a_{i_{mm}}^t$, $a_{i_{me}}^t$, and $a_{i_{THz}}^t$ to indicate whether flow $i$ is scheduled
in the
28GHz band,
E-band, and THz band in slot $t$, respectively. If flow $i$ is scheduled in 28GHz band
in time slot $t$, we have
$a_{i_{mm}}^t=1$; otherwise, $a_{i_{mm}}^t=0$. $a_{i_{me}}^t$ and $a_{i_{THz}}^t$ are defined similarly.

Due to
the HD transmission mode,
the flows that share the same node as their transmitters or receivers are conflicting with each other, and they cannot be scheduled concurrently in any transmission band. If there are conflicts between flows
$i$ and $j$, they
are considered as adjacent flows in
a conflict graph model (where each vertex represents a flow and an edge indicates conflict between the two flows).
We denote conflicting flows $i$ and $j$ as \cite{conf}
\begin{equation}\label{adjacent}
i\propto j, \; \text{if there is a conflict between flows $i$ and $j$}.
\end{equation}
For adjacent flows $i$ and $j$, we can obtain the following constraint.
\begin{equation}\label{con1}
a_{i_{\alpha}}^t+a_{j_{\beta}}^t\leq 1, \alpha,\beta\in\{\text{mm, me, THz}\}.
\end{equation}

Each flow $i$ can only be scheduled in one transmission band,
leading to the following constraint.
\begin{equation}\label{con2}
a_{i_{mm}}^t+a_{i_{me}}^t+a_{i_{THz}}^t\leq 1.
\end{equation}

In the
real heterogenenous network scenario,
different service flows usually
have different QoS requirements. We assume the minimum QoS requirement of flow $i$ is denoted by $q_i$.
Moreover,
the actual transmission rate of flow $i$,
changes over time, because the concurrent transmission flows of flow $i$ in the same band may be different in different TSs.
The transmission rate of flow $i$ at TS $t$ is denoted by $R_i^t$.
According to the transmit band of flow $i$, $R_i^t$ can be calculated by~(\ref{rm}) or~(\ref{rt}). For flow $i$, the actual throughput $T_i$ of a frame is
\begin{equation}\label{con3}
T_i=\frac{\sum_{t=1}^M R_i^t\cdot\Delta t }{t_0+M\cdot\Delta t},
\end{equation}
where $\Delta t$ is the time duration of one TS, and $t_0$ is the time duration of the scheduling phase. And then the constraint of $Q_i$ can be
formulated as
\begin{equation}\label{con4}
Q_i=
\left\{ \begin{array}{ll}
1, & \text{if} \ T_i\geq q_i\\
0, & \text{otherwise}.
\end{array} \right.
\end{equation}

In summary, the problem of optimal scheduling can be formulated as follows.
\begin{equation}\nonumber
\begin{split}
\text{max}  & \;\; \sum_{i=1}^F Q_i\\
\text{s.t.} & \;\;
(\ref{con1})-(\ref{con4}).
\end{split}
\end{equation}
This optimal scheduling problem is a nonlinear integer programming problem, and is NP-hard~\cite{problem}. Consequently, a heuristic algorithm which has
low complexity
is needed
to solve this NP-hard problem in practice.

\section{ Triple-band QoS-based Scheduling Algorithm}\label{S5}

In this section, we propose the QoS-aware
scheduling algorithm
for the three transmission bands, i.e.,the 28GHz
band, the E-band,
and the THz
band. For the proposed algorithm, we first choose the appropriate transmission band for each flow, and then schedule flows to maximize the number of flows with their QoS requirements satisfied.

\subsection{The Transmission Band Selection Algorithm}\label{S5-1}
\begin{algorithm}[tp!]
	\caption {Transmission Band Selection Algorithm} \label{alg:selection algorithm}
	\KwIn {the location of each BS,
	the set of
	flows $S_{all}$ and the QoS requirement of each flow, set $S_{mm}=\emptyset$, $S_{me}=\emptyset$, $S_{THz}=\emptyset$}
    \KwOut{$S_{mm}$, $S_{me}$, $S_{THz}$}
	Calculate $Mq^{mm}_i$, $Mq^{me}_i$ and $Mq^{THz}_i$ for each flow
	Remove $\mathbb{D}=\{i|d_{s_ir_i}> D_{ref}^{THz}\& q_i>Mq^{me}_i\}$ from $S_{all}$;
		\If{$|S_{all}|\neq 0$}{
		\For{flow $i$ ($1\leq i\leq |S_{all}|$)}{
		\If{$d_{s_ir_i}> D_{ref}^{THz}$}{
		\If{$q_i>Mq^{mm}_i$}{
		$S_{me}=S_{me}\cup i$;}
		\Else{
		\If{$C_{me}(i)>C_{mm}(i)$}{
		$S_{mm}=S_{mm}\cup i$;}
		\Else{
		$S_{me}=S_{me}\cup i$;}
        }
		}
		\Else{
		\If{$q_i>Mq^{me}_i$}{
		$S_{THz}=S_{THz}\cup i$;}
		\Else{
		\If{$q_i>Mq^{mm}_i$}{
		\If{$C_{me}(i)\geq C_{THz}(i)$}{
	     $S_{THz}=S_{THz}\cup i$;}
		\Else{
		 $S_{me}=S_{me}\cup i$;}
		}
		\Else{
		 $CM=\text{min}\{C_{mm}(i), C_{me}(i), C_{THz}(i)\}$;\\
		\If{$C_{\omega}(i)=CM$, $(\omega\in\{\text{mm,me,THz}\})$}{
		$S_{\omega}=S_{\omega}\cup i,(\omega\in\{\text{mm,me,THz}\})$;}
        }
		}
        }
		}}
\end{algorithm}

The transmission band selection is based on the QoS requirements and the distances between the transmitters and receivers of the
flows. We first determine the flows that need to be scheduled in each band based on the transmitter-receiver ranges and the flows QoS requirements.
We then schedule the flows in each time slot for transmission.

THz band communications can provide the rates up to multi-Gbps, but its propagation loss is much higher than mmWave communications. Therefore, the THz band is used for only short-distances.
We assume each BS is able to transmit at 28GHz band, E-band, and THz band,
and it can switch among
three bands for different flows. The intelligent switching first
considers
the distance between the transmitter and receiver of each flow. Thus, a reference distance $D_{ref}^{THz}$ is set, which indicate the maximum communication range of THz communications. For the reference distance of mmWave communications, there is
no limit. The communication distances in this paper can satisfy the distance requirement of mmWave communications.
For flow $i$, the choice
between mmWave and THz bands can be decided as follows.
\begin{equation}\label{switch}
switch(i)=
\left\{
\begin{aligned}
&\text{THz} & & \text{if } 0<d_{s_ir_i}\leq D_{ref}^{THz}\\
&\text{mmWave} & & \text{if }  d_{s_ir_i}>D_{ref}^{THz}.
\end{aligned}
\right.
\end{equation}

In addition to the distance, the QoS requirement is also an important basis for each flow's choice of transmission band. Thus, we explore the transmission capability of each band for further flow scheduling. According to~(\ref{rm}) and~(\ref{rt}),
we calculate the transmission rates of flows in different bands, i.e., $R_i^{mm}$ (28GHz mmWave band), $R_i^{me}$ (E-band) and $R_i^{THz}$. In this process, different rates are obtained based on the maximum bandwidth of different bands, and we do not consider the interference from other flows in this calculation. Then the maximum throughput allowed for each band can be expressed as
\begin{equation}\label{capability}
Mq^{\delta}_i=\frac{M\cdot\Delta t\cdot R_i^{\delta}}{t_0+M\cdot\Delta t}, \; \delta\in\{\text{mm, me, THz}\}.
\end{equation}

Therefore, we can draw conclusions as follows:
\begin{enumerate}
\item If $q_i<Mq^{mm}_i$, $i$ can be scheduled in any band.
\item If $Mq^{mm}_i<q_i<Mq^{me}_i$, $i$ can be scheduled in E-band or THz band.
\item If $Mq^{me}_i<q_i<Mq^{THz}_i$, $i$ can only be scheduled in THz band.
\item If $Mq^{THz}_i<q_i$, $i$ is abandoned, and is not considered in the later scheduling process.
\end{enumerate}
Note that
$Mq^{\delta}_i$ is larger than the actual maximum QoS requirement allowed for each band, because the interference from other flows is ignored during the rate calculation.

According to the transmission distances and QoS requirements, one flow may have multiple bands which can meet its transmission conditions. In this case, the flow needs to further select among multiple bands. We indicate the sets of the flows in 28GHz band, E-band and THz band as $S_{mm}$, $S_{me}$ and $S_{THz}$, respectively. Then we define a parameter to compare the adaptability of each band with respect to flow transmissions. In order to schedule more flows in one frame, it is important to minimize the number of conflicting flows in each band.
Hence, we define this comparison parameter of a band as the sum of the number of slots required by conflicting flows in this band. For flow $i$, the comparison parameter of the $\omega$ ($\omega\in\{\mbox{mm, me, THz}\}$) band can be defined as
\begin{equation}\label{comparison}
C_\omega(i)=\sum_{\{i^\prime|i^\prime\in S_\omega,i\propto i^\prime\}}{\frac{q_{i^\prime}}{R_{i^\prime}^{\omega}}}, \omega\in\{\text{mm, me, THz}\},
\end{equation}

The backhaul scheduling algorithm is summarized in Algorithm~\ref{alg:selection algorithm}.
It works as follows.
First, remove
flows which do not satisfy the transmission conditions of these three bands in Line 1. In addition to the flows in set $\mathbb{D}$, other flows are divided into two categories based on whether their transmission distances are shorter than the maximum transmission
range
of the THz band in Lines 4-31.
In Lines 5-6, some flows satisfying
$d_{s_ir_i}> D_{ref}^{THz}$
will be put in
the E-band since
their QoS requirements are
larger than the maximum throughput allowed in the 28GHz band. In
Lines 8-11,
other flows that satisfy $d_{s_ir_i}> D_{ref}^{THz}$ and $q_i<Mq^{mm}_i$
will be put in either the 28GHz band or the E-band
according to the comparison parameter
defined in~(\ref{comparison}).
In Lines 15-30,
flows that satisfy $d_{s_ir_i}\leq D_{ref}^{THz}$ can be divided into three categories according to their QoS requirements,
i.e., flows that can only
be transmitted in the THz band, flows that can
be transmitted
in both THz band and E-band,
and flows that can
be transmitted
in all the
three bands. Flows are put into
the THz band and E-band according to the comparison parameter~(\ref{comparison}) in Lines 18-23.
Finally,
flows that can
be transmitted
in all the
three bands
will be scheduled in one of the bands
according to the comparison parameter in
Lines 25-27.

For the complexity of the selection algorithm,
the
number of the iterations of the outer {\tt for} loop in Line 3 is $|S_{all}|$, where $|S_{all}|$ in the worst case is $\mathcal{O}(F)$. So the computational complexity of this algorithm is $\mathcal{O}(F)$.

\subsection{ The Backhaul Scheduling Algorithm }\label{S5-2}

\begin{algorithm} [tp!]
	\caption {Triple-Band Backhaul Scheduling Algorithm} \label{alg:scheduling algorithm}
	\KwIn{ $S_{mm}$, $S_{me}$, $S_{THz}$, the location of each 
	BS
	and the QoS requirement of each flow, the
	scheduled flow
	set $Sch=\emptyset$, the
	$F\ast M$ scheduling matrix $\mathbf{A=0}$, the number of completed flows $N_{com}=0$}
    \KwOut{$\mathbf{A}$, $N_{com}$}
        Calculate the degrees and the priority values of the flows in each band;\\
		\For{slot $t$ $(1\leq t\leq M)$}{
		 $S=S_{mm}\cup S_{me}\cup S_{THz}$;\\
		 Consider the degree first and rearrange the flows in $S$ by increasing degree, then sort the flows with the same degree in descending order by priority value;\\
		\If{$|S|\neq 0$}{
		\For{each flow $i\in S$}{
		Find the set $S_{\gamma}$ that $i\in S_{\gamma}$ $(\gamma\in\{\text{mm, me, THz}\})$;\\
		\If{$i$ has no contention with the flow(s) in $Sch$ }{
		$Sch=Sch\cup i$;\\
		 $S=S-i$;\\
		 $S_{\gamma}=S_{\gamma}-i$ $(\gamma\in\{\text{mm, me, THz}\})$;}
		}
		\For{each flow $j\in Sch$}{
		 $\mathbf{A}(j,t)=1$;\\
		 Calculate the rate $R_j^\delta$ $(\delta\in\{\text{mm, me, THz}\})$ in the current time slot;\\
		 Calculate the remaining throughput demand of Flow $j$; \\
         $q_j=q_j\ast(t_0+M\cdot\Delta t)-R_j^\delta\ast\Delta t$;\\
		\If{$q_j\leq 0$}{
		 $Sch=Sch-j$;\\
		 $Q_j=1$;\\
		 $N_{com}=N_{com}+1$;}
		}	}
        \Else{break}
		}
\end{algorithm}

In this section, we propose the QoS-based scheduling algorithm which can schedule as many flows as possible
in
each band. Actually, the scheduling algorithm based on the flow QoS requirements
is similar to the time division multiplexing algorithm (TDMA). We divide time into multiple TSs and schedule the appropriate flows in each TS. Because of the multi-band transmission, more flows can be scheduled simultaneously
while
achieving their throughput requirements. To clearly present the algorithm, we first introduce the conflict limit between flows and the priority of flow scheduling.

Regarding
the contention
among
flows within
each band, we consider two cases.
First, the flows
share the same BS
as
their transmitter or receiver
cannot
transmit data at the same time, because of the HD
mode.
Second,
the interference on the flow from another flow is so severe,
such that these two flows
cannot
be scheduled concurrently. For the second case, we define a parameter to represent the relative interference between flows as follows.
\begin{equation}\label{RI}
RI^\delta(j,i)=\frac{I^\delta_{ji}}{P_r^\delta(i,i)}, \; \delta\in\{\text{mm, me, THz}\},
\end{equation}
where $I^{mm}_{ji}$ and $I^{me}_{ji}$ are defined by~(\ref{im}), and $I^{THz}_{ji}$ is defined by~(\ref{it}). $P_r^{mm}(i,i)$ and $P_r^{me}(i,i)$ are defined by~(\ref{pm}), and $P_r^{THz}(i,i)$ is defined by~(\ref{pt}). For the
mutual
interference
among
the flows
that are transmitted
concurrently, thresholds $\sigma^{mm}$, $\sigma^{mm}$, $\sigma^{THz}$ are given for 28GHz band, E-band and THz band, respectively. Hence, we draw the following conclusion
\begin{equation}\label{concul}
\begin{split}
a_{i_{\delta}}^t+a_{j_{\delta}}^t\leq 1, \text{if}\, RI^\delta(j,i)>\sigma^\delta,
\delta\in\{\text{mm, me, THz}\},
\end{split}
\end{equation}

In addition to conditions for concurrent transmissions, the order of the flow scheduling also needs to be determined. We define the number of flows which share the same BSs
with flow $i$ as the {\em degree} of flow $i$, and schedule flows with
a smaller degree preferentially.
This way,
more flows can be scheduled
to be transmitted
simultaneously. If the degrees of multiple flows are the same, we define another parameter to prioritize the scheduling order of the flows,
which is
the inverse of a flow's
required number of TSs in a frame
to satisfy
its QoS requirement. Flows that can reach QoS requirements quickly can complete their transmissions soon and leave time resources to other flows. Therefore, prioritizing flows that require less time slots can transmit more flows in the fixed time.
For
flow $i$, the priority value can be expressed as
\begin{equation}\label{priority}
pri(i)=\frac{R_i^\delta\cdot\Delta t }{q_i\ast(t_0+M\cdot\Delta t)}, \; \delta\in\{\text{mm, me, THz}\}.
\end{equation}
In~(\ref{priority}),
the transmission rate $R_i^\delta$ is not the actual rate;
it is the estimate value without considering the interference from other flows. For each band, we calculate the priority value of each flow in the corresponding set according to~(\ref{priority}), and schedule the corresponding flows in the descending order in the subsequent process.

For the QoS-based backhaul scheduling, flows that need to be scheduled in each frequency band are obtained first, and then we record the flows of three frequency bands as three
non-overlapping
sets, i.e., $S_{mm}$, $S_{me}$, and $S_{THz}$
for the flows in
the
28GHz band, E-band and THz band, respectively, and  $|S_{mm}|+|S_{me}|+|S_{THz}|\leq F$.
We denote the scheduling scheme by an
$F\ast M$ binary matrix $\mathbf{A}$, where $\mathbf{A}(i,t)=1$ means
flow $i$ is scheduled in slot $t$.
We also
propose a parameter $N_{com}$ to indicate the number of
completed flows.

When scheduling flows, the scheduling algorithm selects flows that can transmit simultaneously from each set by
their
priority. Selecting the flows
having
no contention is based on the two principles mentioned earlier: (i) concurrent flows cannot share the same
BS to be their transmitter or receiver;
(ii) the
mutual interference between concurrent flows in the same band cannot exceeded the threshold.
If flows that wait for being scheduled follow the above two principles, they can be concurrently transmitted with flows being transferred.
If a
flow
is scheduled with its QoS requirement satisfied,
it will
no longer be
assigned any time slot and another new flow will be selected for concurrent transmission with other current transmitting flows. This way can prevent the waste of time slots and
serve more flows.

The backhaul scheduling algorithm is summarized in Algorithm~\ref{alg:scheduling algorithm}.
After the
initialization steps,
we rearrange flows by increasing degree and decreasing priority value in Lines 2-3. In Lines 4-13, we find new flows
that
can be concurrently transmitted with
the
existing, scheduled flows. In Lines
14-23,
we calculate the rate and remaining throughput demand of each
scheduled
flow, and set the scheduling matrix. In the end we obtain the
flow transmission schedule
and the number of completed flows.

For the complexity of the scheduling algorithm, we can see that the outer {\tt for} loop has $\mathcal{O}(M)$ iterations. The {\tt for} loop in
Line 5 has $|S|$ iterations, and $|S|$ in the worst case is $\mathcal{O}(F)$. Besides, the loop of contention verification in
Line 7 has $|Sch|$ iterations, and $|Sch|$ in the worst case is $\mathcal{O}(F-1)$. Therefore, the complexity of the triple-band backhaul scheduling algorithm is $\mathcal{O}(MF(F-1))$. The low complexity of the scheduling algorithm reflects its
value for practical implementations.

\section{Performance Evaluation}\label{S6}

In this section, we evaluate the performance
of the triple-band transmission backhaul scheduling algorithm. The algorithm involves three bands
for flow transmissions: the 28GHz band, the 73GHz band (i.e., the E-band),
and the
340GHz (i.e., the THz band).
We also compare the performance of the proposed algorithm with several baseline schemes.

\subsection{Simulation Setup}\label{S6-1}

We consider a backhaul network
deployed within
a $100m\times 100m$ square area. There are 20 BSs which are randomly distributed in this area and at most 350 flows need to be scheduled. Each flow is generated with randomly selecting its source and destination among all BSs. The QoS requirement of each flow is uniformly distributed between 1Mbps and 10Gbps. Other parameters are shown in Table~\ref{tab:parameters}.

The transmission powers of mmWave communication and THz communication are about two orders of magnitude different in this paper, so we set the mutual interference thresholds $\sigma^{THz}=10^{-2}$ and $\sigma^{mm}=\sigma^{me}=10^{-4}$.

According to the current research, the distance coverage of THz communications can reach about 50m. So we assume the reference distance of THz communications with $D^{THz}_{ref}=50m$. The transmission
range
of THz communications can be extended with
a larger transmitter gain~\cite{THZ}.
However, we do not consider
such cases in the simulations.

\begin{table}[htbp]
\captionsetup{font={small}}
 \caption{\label{tab:parameters} Simulation Parameters}
 \centering
 \begin{tabular}{lll}
  \toprule
  Parameter&Symbol&Value\\
  \midrule
  mmWave transmission power&$P_t^{mmWave}$&1W\\
  THz transmission power&$P_t^{THz}$&20mW\\
  MUI factor&$\rho$, $\rho^\prime$&1\\
  transceiver efficiency factor&$\eta$&0.5\\
  path loss exponent&$n$&2\\
  28GHz bandwidth&$W^{mm}$&800\,MHz\\
  E-band bandwidth&$W^{me}$&1.2\,GHz\\
  THz bandwidth&$W^{THz}$& 10\,GHz\\
  background noise&$N_0$&-134dBm/MHz\\
  minimum antenna gain&$G^{min}$&0 dB\\
  maximum antenna gain&$G^{max}$&20 dB\\
  beamwidth of the main lobe&$\theta_{ml}$&$\pi/6$\\
  slot time&$\Delta t$&$18\mu s$\\
  beacon period&$t_0$&$850\mu s$\\
  number of slots&$M$&$2000$\\
  \bottomrule
 \end{tabular}
\end{table}

For comparison purpose, we implement the common single-band scheme and the dual-band scheme and the state-of-the-art
Maximum QoS-aware Independent Set (MQIS) based scheduling algorithm~\cite{MQIS} as baseline schemes.
\begin{itemize}
  \item Single-band Scheme:
  this is essentially the same as the proposed triple-band scheduling scheme, but it only operates
  in one band. Considering the distance limitation of the
  THz band and the bandwidth limitation of the lower-frequency mmWave band, E-band is selected
  for this single-band scheme.

  \item Dual-band Scheme:
  this scheme operates
  in 2.4GHz (with 20MHz bandwidth) and 60GHz (with 2.16GHz bandwidth)~\cite{dual}, and the scheduling is similar to
  the proposed scheduling scheme. Currently, the dual-band cooperation of sub-6GHz and mmWave has been extensively studied.

  \item MQIS: this is the concurrent scheduling algorithm based on the maximum QoS-aware independent set proposed in~\cite{MQIS}. We apply triple-band cooperation to the MQIS algorithm. Flows are divided into multiple independent sets. When all flows in one set are scheduled and completed, flows in another set can start to be scheduled.
\end{itemize}

For the evaluation study, we consider the following performance metrics:
\begin{itemize}
  \item Number of completed flows: the number of scheduled flows with their QoS requirements being satisfied. Those flows that have been scheduled but
  their QoS requirements cannot be met
  are not counted as a completed flow.

  \item System throughput: the total throughput that the network
  system can achieve. This metric is the sum
  of the
  average throughputs
  of all flows carried
  in the network.
\end{itemize}

\subsection{Comparison with Existing Schemes} \label{S6-2}

\begin{figure}[!t]
	\begin{center}
		\includegraphics[width=3.3in]{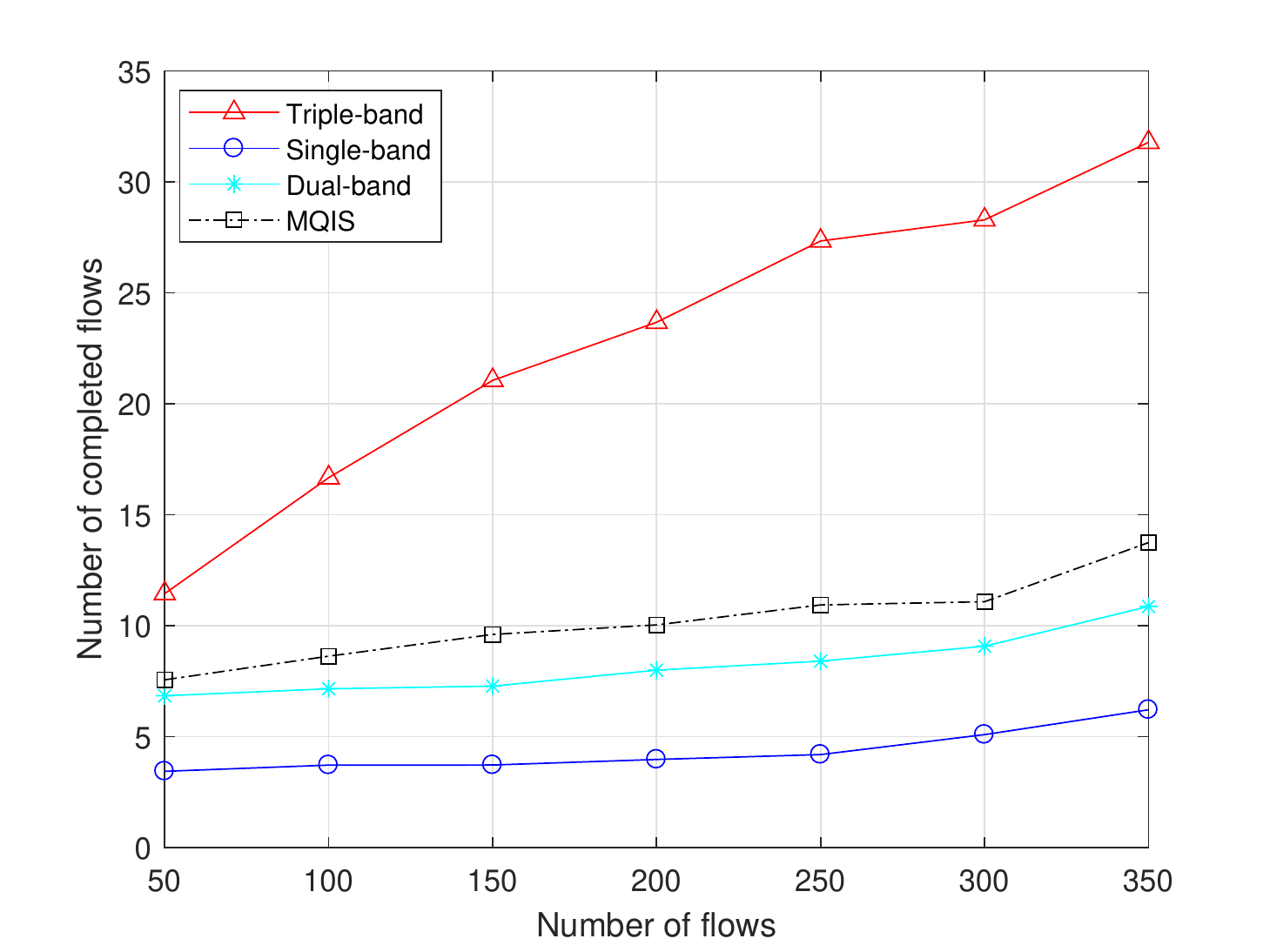}
		\captionsetup{font={small}}
		\caption{Number of completed flows versus different numbers of requested flows.} \label{fig:flow-flow}
	\end{center}
\end{figure}

\begin{figure}[!t]
	\begin{center}
		\includegraphics[width=3.3in]{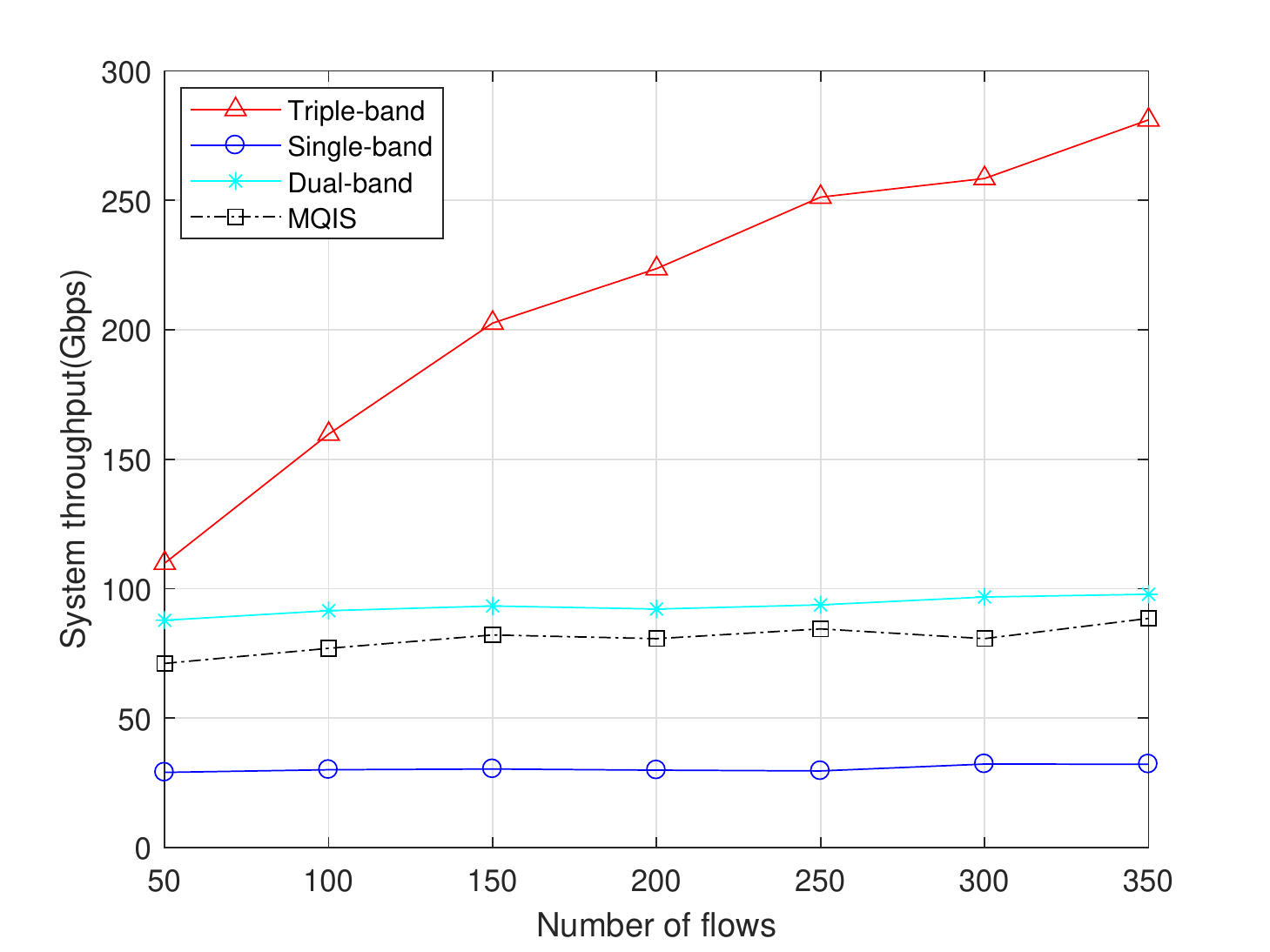}
		\captionsetup{font={small}}
		\caption{System throughput versus different numbers of requested flows.} \label{fig:throughput-flow}
	\end{center}
\vspace{-0.15in}
\end{figure}

In Fig.~\ref{fig:flow-flow} and Fig.~\ref{fig:throughput-flow}, the number of time slots is set to 2000. The abscissas of these two figures are both the number of requested flows, which
is varied
from 50 to 350.
As
the number of requested flows
is increased,
Fig.~\ref{fig:flow-flow} and Fig.~\ref{fig:throughput-flow} plot
the simulation
results of the number of completed flows and system throughput, respectively.

From Fig.~\ref{fig:flow-flow}, we can see that the
trend of the proposed triple-band scheme curve is rising with the
increased number of flows.
The more flows that need to be scheduled,
the more flows can be scheduled simultaneously and the more spatial reuse comes into play. Because of the system capacity limitation, the growth of the number of completed flows of the triple-band scheduling scheme
gets slowed down with the further increasing of the number of offered flows.
Compared with our proposed scheme, although there is also a
trend of
growth
in the number of completed flows of MQIS,
the increase is much slower than that of the proposed scheme.
The ability of MQIS to schedule flows is inferior to the proposed triple-band scheme, because MQIS
cannot
schedule flows without interruption, which
wastes a certain amount of the time resource.
For the dual-band scheme and the single-band scheme, there are also slower
growths
in the number of completed flows along the x-axis direction.
This is because the single-band scheme does not
exploit
the THz band,
and its ability to schedule the flows is thus
limited. Besides, the completed flows of the dual-band scheme and the single-band scheme are both relatively few. Some flows with large QoS requirements cannot be scheduled without the THz band. When the number of flows is 350, the proposed triple-band scheduling scheme improves the number of completed flows by
56.3\% compared with the MQIS scheme, by 64.1\% compared with the dual-band scheme, and by
79.7\% compared with the single-band scheme.

From Fig.~\ref{fig:throughput-flow}, we can observe that the
trend of the throughput curves is similar to the
trend of the number of completed flows. There is a certain gap between the completed-flow numbers of
the four schemes.
and the gaps of throughput curves
are similar as that in Fig.~\ref{fig:flow-flow} except for the dual-band scheme.
For the dual-band scheme, we set a very large bandwidth for one of its frequency bands (even larger than $W^{me}$), so its performance on the system throughput can be better than MQIS. Specifically, when the number of flows is 350, the proposed triple-band scheduling scheme improves the system throughput by 67.9\% compared with the dual-band scheme,
by 64.3\% compared with MQIS, and by
87.5\% compared with the single-band scheme.

\begin{figure}[!t]
	\begin{center}
		\includegraphics[width=3.3in]{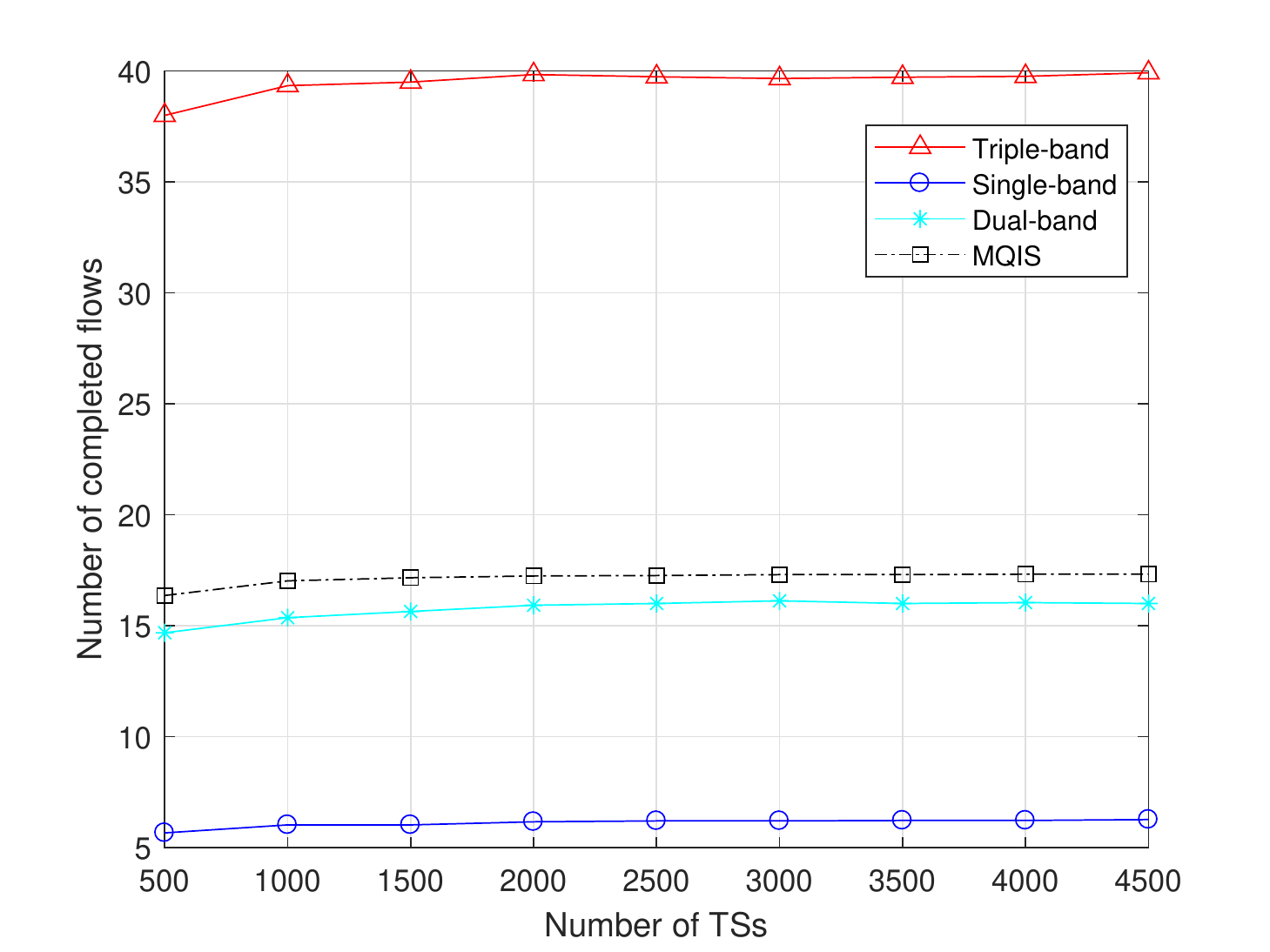}
		\captionsetup{font={small}}
		\caption{Number of completed flows versus different numbers of time slots.} \label{fig:flow-TS}
	\end{center}
\end{figure}

\begin{figure}[!t]
	\begin{center}
		\includegraphics[width=3.3in,height=5.55cm]{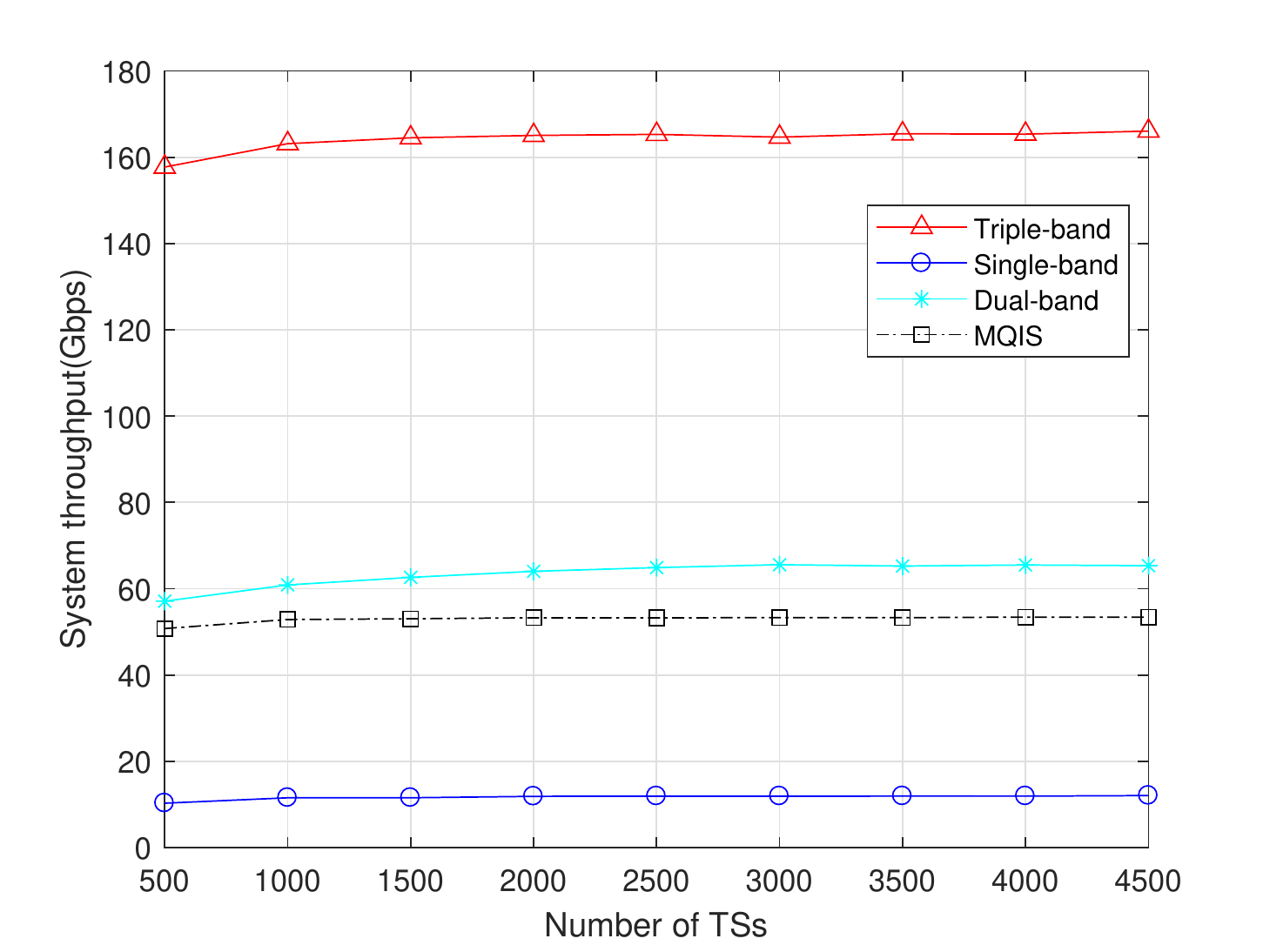}
		\captionsetup{font={small}}
		\caption{System throughput versus different numbers of time slots.} \label{fig:throughput-TS}
	\end{center}
\end{figure}

In Fig.~\ref{fig:flow-TS} and Fig.~\ref{fig:throughput-TS}, the number of flows is
fixed at
350. The number of time slots is
increased from 500 to 4500. With the number of TSs
is increased,
Fig.~\ref{fig:flow-TS} and Fig.~\ref{fig:throughput-TS} plot simulation results of the number of
completed flows and system throughput.

From Fig.~\ref{fig:flow-TS} and Fig.~\ref{fig:throughput-TS}, we can see that the number of completed flows and system throughput have the same trend
with increased
number of TSs. These schemes also
exhibit
similar trends.
When the number of time slots starts to increase, both the number of completed flows and system throughput increase. When the number of time slots
is increased
over
a certain value (
i.e., to 2000 in Fig.~\ref{fig:flow-TS} and Fig.~\ref{fig:throughput-TS}), the number of completed flows and system throughput become saturated and do
not increase significantly anymore.
This shows that properly extending the duration of each frame is conducive to improving the
transmission performances, but the improvement is limited. In this case, the numbers of completed flows of the proposed triple-band scheduling scheme, MQIS, the dual-band scheme and the single-band scheme can be up to 40, 18, 16 and 7, respectively.
Furthermore,
it is not necessary to set a very large value for the number of TSs.

\begin{figure}[!t]
	\begin{center}
		\includegraphics[width=3.3in]{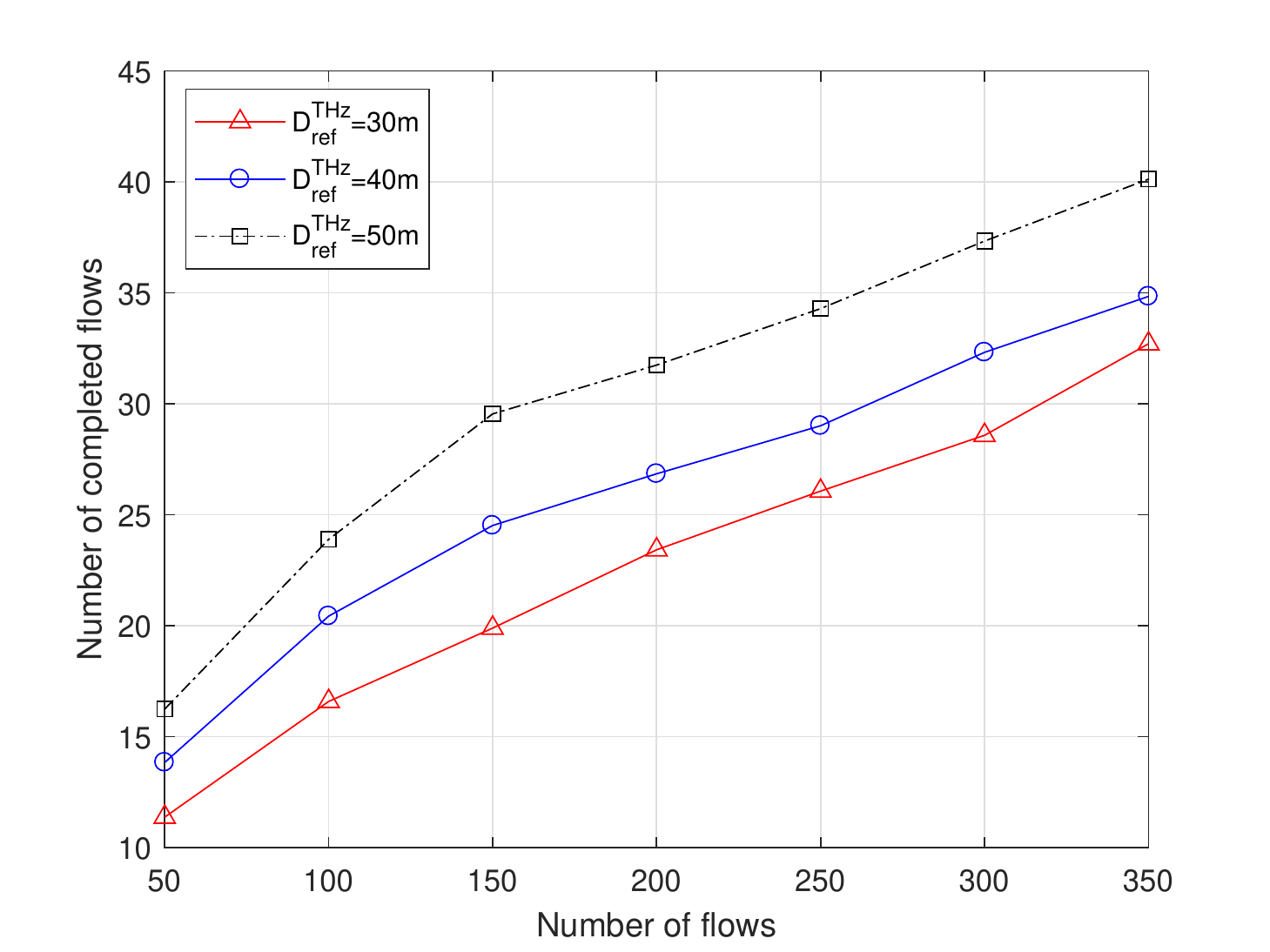}
		\captionsetup{font={small}}
		\caption{Number of completed flows versus different reference distances of THz communications.} \label{fig:flow-D}
	\end{center}
\end{figure}

\begin{figure}[!t]
	\begin{center}
		\includegraphics[width=3.3in,height=5.55cm]{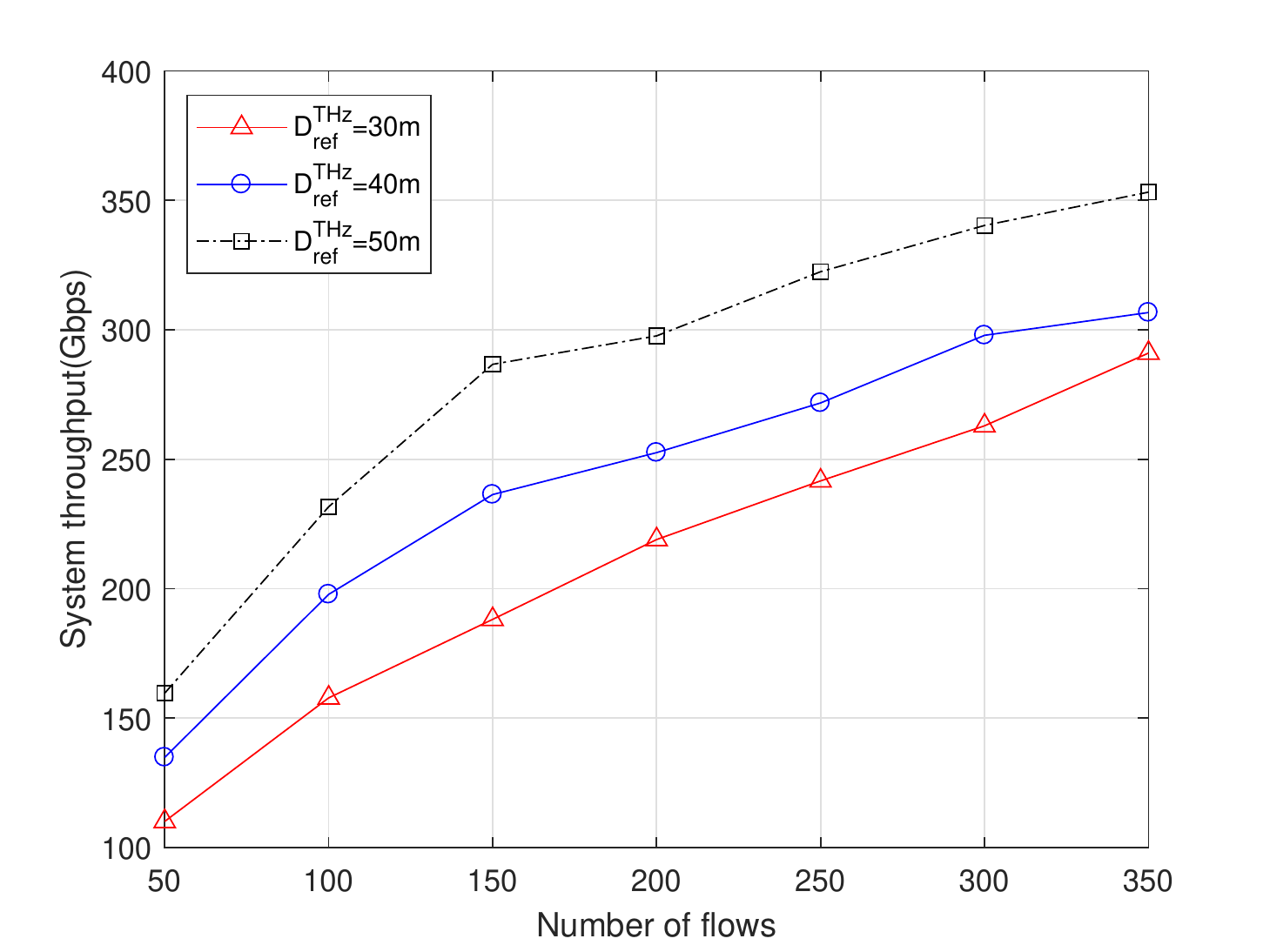}
		\captionsetup{font={small}}
		\caption{System throughput versus different reference distances of THz communications.} \label{fig:throughput-D}
	\end{center}
	\vspace{-0.1in}
\end{figure}

In Fig.~\ref{fig:flow-D} and Fig.~\ref{fig:throughput-D}, the number of flows is set to 350 and the number of time slots is set to 2000. We evaluate the impact of the reference distance of THz communications, i.e., $D_{ref}^{THz}$, which is set to 30m, 40m, and
50m. With
increased number of
flows, Fig.~\ref{fig:flow-D} and Fig.~\ref{fig:throughput-D} plot the number of completed flows and system throughput under different $D_{ref}^{THz}$ values.

From Fig.~\ref{fig:flow-D} and Fig.~\ref{fig:throughput-D}, we can see that the number of completed flows and system throughput
exhibit similar results
as
$D_{ref}^{THz}$ is increased.
The larger the $D_{ref}^{THz}$,
the more flows can be scheduled simultaneously and the larger
the
system throughput.
If the reference distance $D_{ref}^{THz}$ is large, there will be more flows
that can be scheduled in the THz band,
leading
to more completed flows and larger throughput. We can also observe that the gap between the curves of  $D_{ref}^{THz}=40$m and $D_{ref}^{THz}=30$m is bigger than the gap between the curves of $D_{ref}^{THz}=50$m and $D_{ref}^{THz}=40$m. The performance improvement decreases with
further increased
$D_{ref}^{THz}$, which shows that the performance improvement caused by $D_{ref}^{THz}$ is also
limited. However, we need to consider the actual transmission characteristics of the THz band,
while
$D_{ref}^{THz}$ cannot be too large. If the distance
of the THz communications
is too large,
the higher propagation loss will cause
more flow transmission to fail.

\begin{figure}[!t]
	\begin{center}
		\includegraphics[width=3.3in,height=6.3cm]{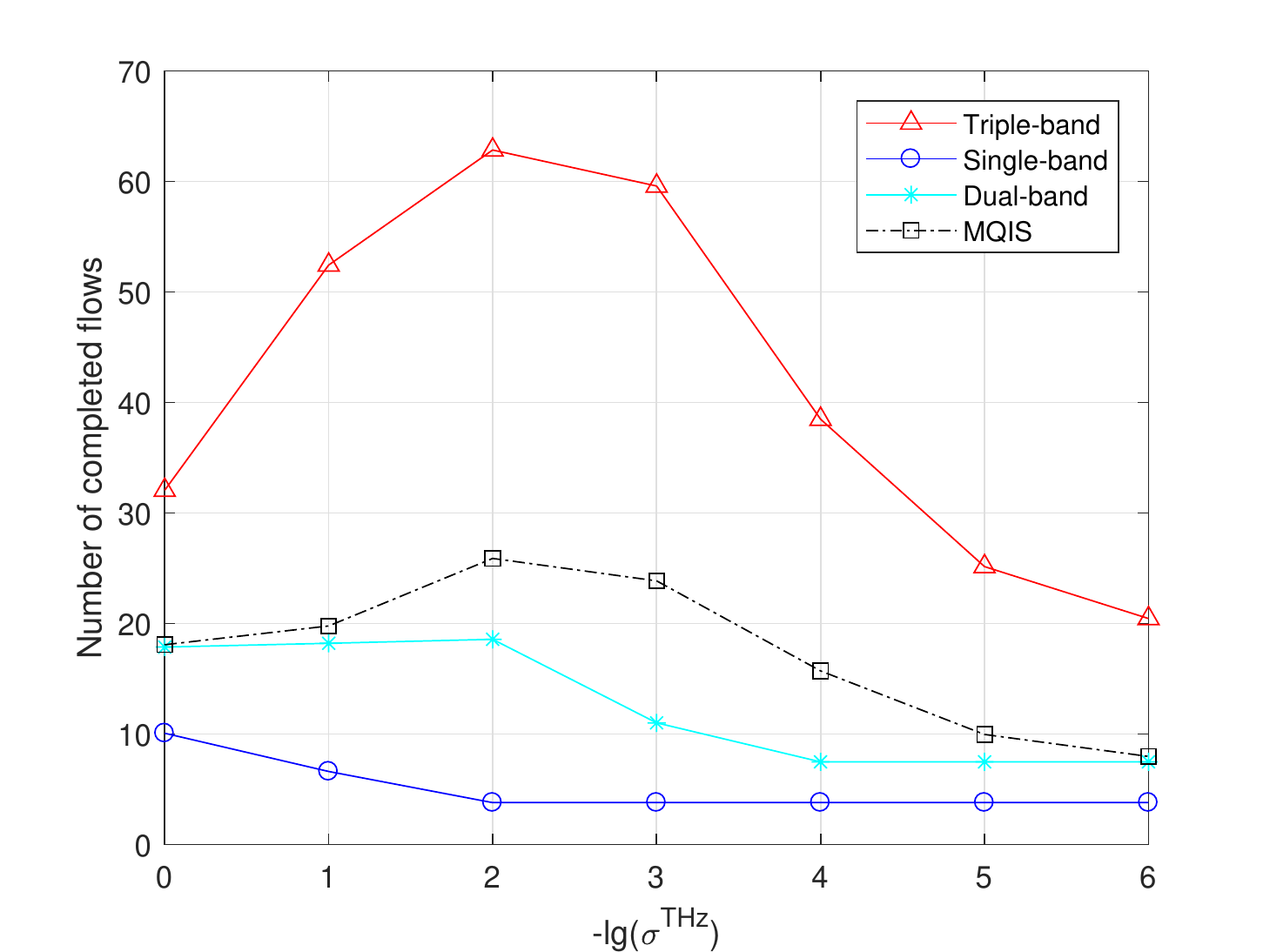}
		\captionsetup{font={small}}
		\caption{Number of completed flows versus different thresholds.} \label{fig:flow-threshold}
	\end{center}
\vspace{0.15in}
\end{figure}

\begin{figure}[!t]
	\begin{center}
		\includegraphics[width=3.3in,height=6.3cm]{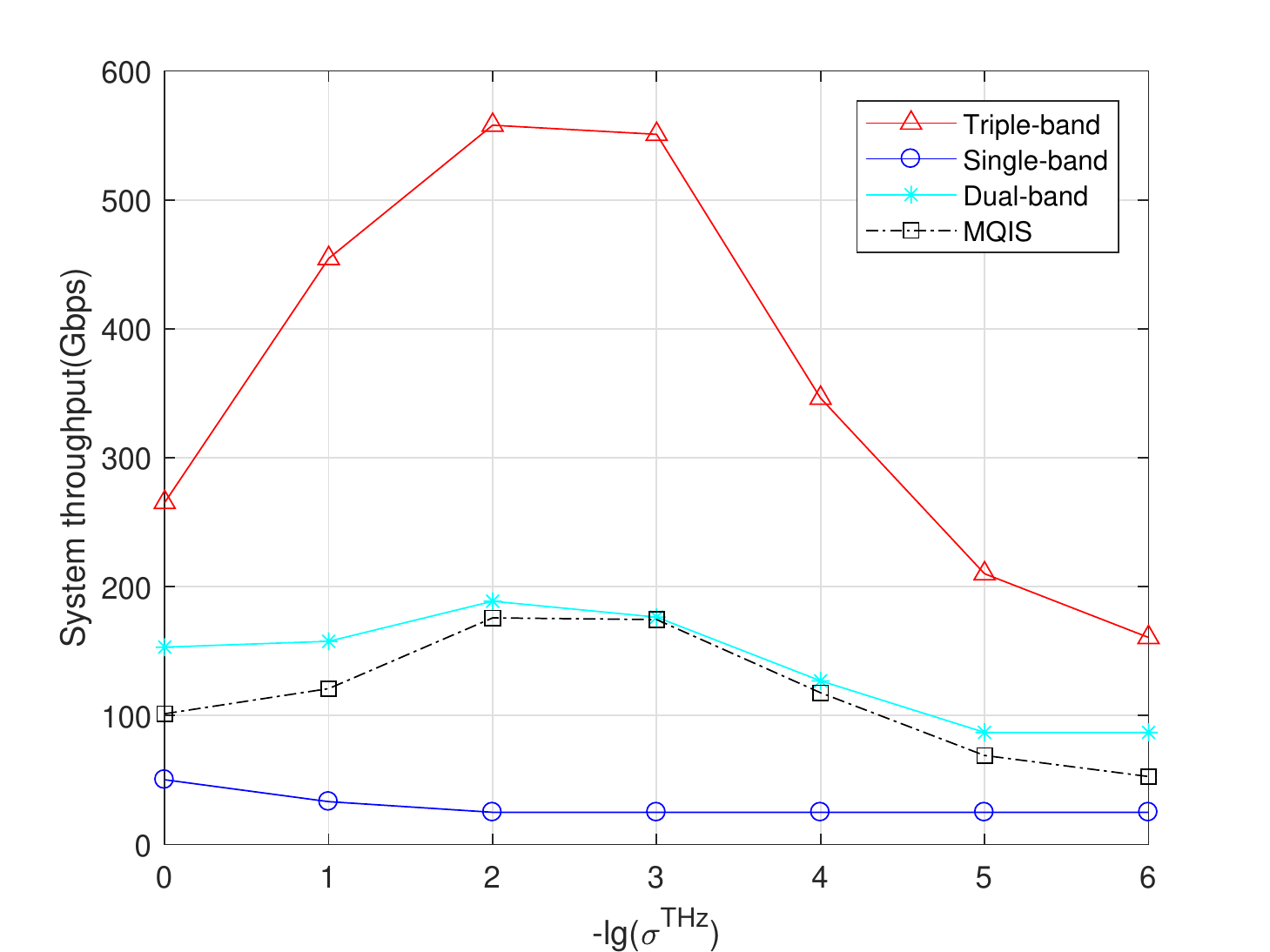}
		\captionsetup{font={small}}
		\caption{System throughput versus different thresholds.} \label{fig:throughput-threshold}
	\end{center}
	\vspace{0.1in}
\end{figure}

In Fig.~\ref{fig:flow-threshold} and Fig.~\ref{fig:throughput-threshold}, the number of flows is set to 350 and the number of time slots is set to 2000. Abscissas of these two figures are
$-\lg(\sigma^{THz})$,
which is
varied
from 0 to 6. $-\lg(\sigma^{mm})$ and $-\lg(\sigma^{me})$ are about two orders of magnitude lower than $-\lg(\sigma^{THz})$, so they are varied from 2 to 8 with $-\lg(\sigma^{THz})$ is varied from 0 to 6.
With decreased threshold, Fig.~\ref{fig:flow-threshold} and Fig.~\ref{fig:throughput-threshold}
plot the
number of
completed flows and system throughput
achieved by the four schemes.

From Fig.~\ref{fig:flow-threshold} and Fig.~\ref{fig:throughput-threshold}, we can
see
that trends of
the number of
completed flows and system throughput are
similar
with
the decrease of interference threshold. For the triple-band scheme, when thresholds start to decrease, both the number of completed flows and system throughput increase.
Properly
decreasings of thresholds can avoid simultaneous scheduling of some flows that have serious interference between themselves. And then their transmission rates can
be
increased, which will lead
to
more flows to be completed more
quickly. When $\sigma^{THz}$, $\sigma^{me}$ and $\sigma^{mm}$ are
decreased
to a certain order of magnitude (
i.e., $\sigma^{THz}=10^{-2}$, $\sigma^{me}=\sigma^{mm}=10^{-4}$ in Fig.~\ref{fig:flow-TS} and Fig.~\ref{fig:throughput-TS}), the number of completed flows and system throughput start to
decrease significantly. A too small threshold will weaken the advantages of spatial reuse and reduce flows that can be scheduled simultaneously. Therefore, the appropriate actual interference threshold is important for the
transmission performance. In the case of
our simulation study,
actual interference thresholds $\sigma^{THz}$ of $10^{-2}$ order, $\sigma^{me}$ and $\sigma^{mm}$ of $10^{-4}$ order
seem to be
suitable.
For the single-band scheme, their scheduling capabilities have been limited by
their specific designs
when the number of flows is 350. Hence, the number of completed flows and system throughput of this scheme decrease with
decreased threshold.

\begin{figure}[!t]
	\begin{center}
		\includegraphics[width=3.3in,height=6.3cm]{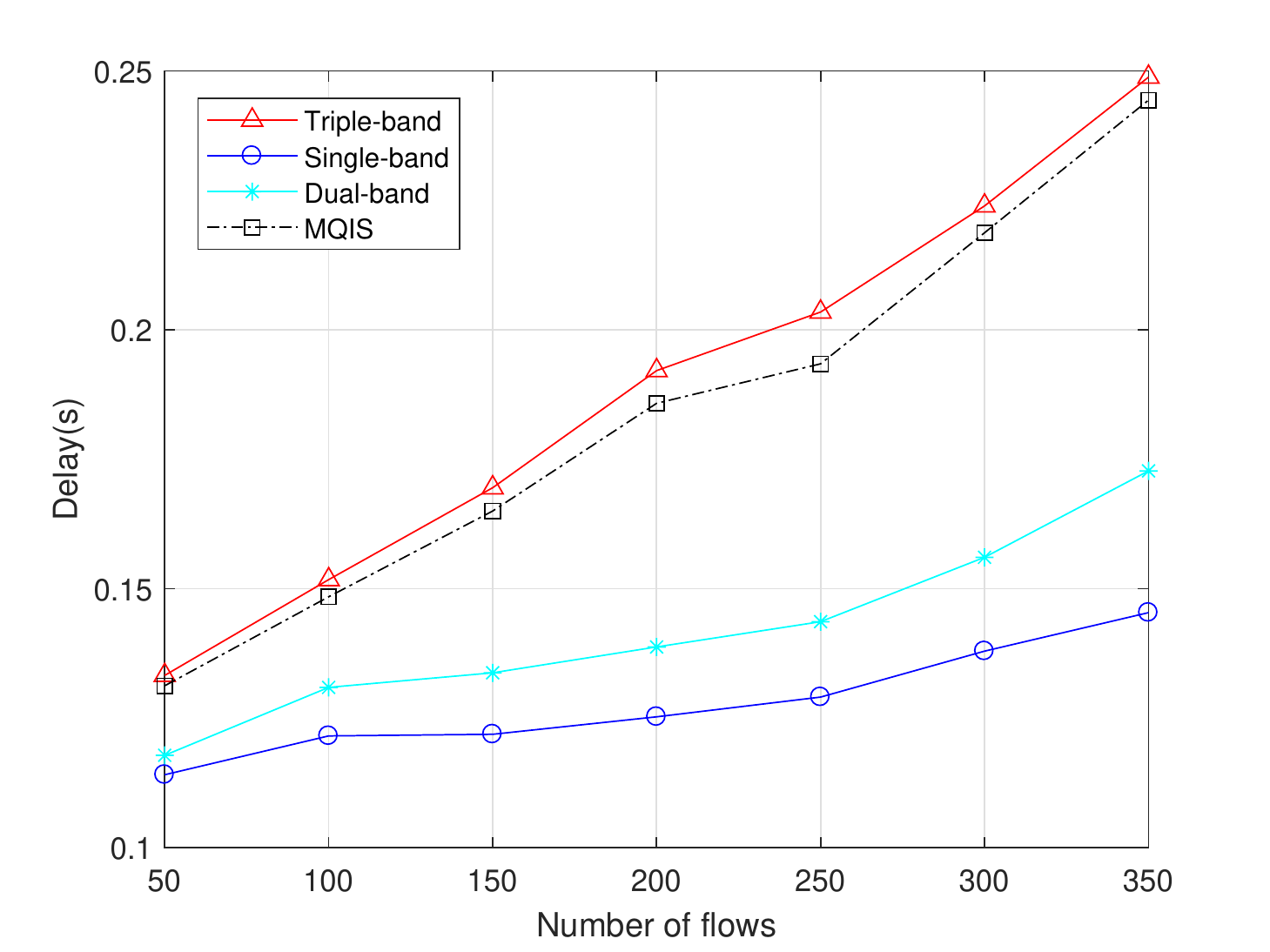}
		\captionsetup{font={small}}
		\caption{Delay versus different numbers of requested flows.} \label{fig:delay}
	\end{center}
	\vspace{0.1in}
\end{figure}
In Fig.~\ref{fig:delay}, the number of time slots is set to 2000. The abscissas of the figure is the number of requested flows, which
is varied
from 50 to 350.
As
the number of requested flows
is increased,
Fig.~\ref{fig:delay} plots
the simulation
results of the delay of different schemes.

From Fig.~\ref{fig:delay}, we can see that the
trends of all the schemes curve are rising with the
increased number of flows.
Compared with our proposed scheme, other schemes have different advantages in terms of delay. However, combined with the advantage of the proposed triple-band scheme shown in Fig.~\ref{fig:flow-flow}, the advantages of other schemes in terms of delay are of little significance. It is obvious that the more scheduled flows, the greater the delay of the scheme.
When the number of flows is 350, the delay of the proposed triple-band scheduling scheme is 0.25s, and it is the delay size that allows communications between BSs.

\section{Conclusions}\label{S7} 

In this paper, we considered
the
problem of scheduling a large number of flows with diverse QoS requirements over three frequency bands (i.e., the 28GHz band, E-band, and THz band).
To maximize the number of flows
while
satisfying
their QoS requirements,
we proposed the triple-band scheduling scheme, which can schedule flows to
be concurrently transmitted
and reduce the waste of resource, while considering the QoS requirements and transmission ranges of the flows.
Extensive simulations showed that our proposed scheme
outperformed three baseline schemes
on the number of
completed flows and system throughput.
However, the complexity of THz transmission itself and the high cost of transmission equipment limit the communication, and we require further research and follow-up. In the future work, we will consider a more realistic scene and transmission model. Besides, we will do the performance verification of our algorithm on the actual system platform to further demonstrate the practicality of our scheme.

\vfill
\end{document}